\documentclass[prb,twocolumn,superscriptaddress,showpacs,floatfix,amsmath,amssymb]{revtex4-1}
\usepackage{dcolumn}
\usepackage{xcolor}
\usepackage[utf8]{inputenc}
\usepackage{amsmath}
\usepackage{amssymb}
\usepackage{dsfont}
\usepackage{physics}
\usepackage{color}
\usepackage{graphicx}
\usepackage{mathtools}
\usepackage{overpic}
\usepackage[toc,page]{appendix}
\usepackage{chemfig}
\usepackage{ccycle}
\usepackage{comment}
\usepackage{textcomp}
\usepackage{gensymb}
\usepackage{graphics}
\usepackage{epsfig}
\usepackage{subfigure}
\usepackage[T1]{fontenc}

\begin{document}
\preprint{APS}

\title{Lattice dynamics of photoexcited insulators from\\ constrained density-functional perturbation theory}

\author{Giovanni Marini}
\affiliation{Graphene Labs, Fondazione Istituto Italiano di Tecnologia, Via Morego, I-16163 Genova, Italy}

\author{Matteo Calandra} 
\affiliation{Graphene Labs, Fondazione Istituto Italiano di Tecnologia, Via Morego, I-16163 Genova, Italy}
\affiliation{Department of Physics, University of Trento, Via Sommarive 14, 38123 Povo, Italy}
\begin{abstract}
We present a constrained density functional perturbation theory scheme for the calculation of structural and harmonic vibrational properties of insulators in the presence of an
excited and thermalized electron-hole plasma. The method is ideal to tame ultrafast light induced structural transitions in the regime where the photocarriers thermalize faster than the lattice, the electron-hole recombination time is longer than the phonon period and the photocarrier concentration is large enough to be approximated by an electron-hole plasma.
The complete derivation presented here includes total energy, forces and stress tensor, variable cell structural optimization, harmonic vibrational properties and the electron-phonon interaction. 
We discuss in detail the case of zone center optical phonons not conserving the number of electrons and inducing a Fermi shift in the photo-electron and hole distributions. We validate our implementation
by comparing with finite differences in Te and VSe$_2$. 
By calculating the evolution of the phonon spectrum of Te, Si and GaAs as a function 
of the fluence of the incoming laser light, we demonstrate that even at low fluences, corresponding 
to approximately $0.1$ photocarriers per cell, the phonon
spectrum is substantially modified with respect to the ground state one with new Kohn anomalies appearing and a substantial softening of zone center optical phonons.  Our implementation can be efficiently used to detect reversible transient phases and irreversible structural transition induced by ultrafast light absorption.

\end{abstract}

\maketitle
\section{Introduction}

The development of Density Functional Perturbation Theory (DFPT) approaches \cite{PhysRevB.43.7231,PhysRevLett.58.1861,PhysRevB.51.6773,RevModPhys.73.515,PhysRevB.54.16487,PhysRevB.55.10355}
has allowed for efficient calculations of the ground-state linear response properties of materials without
need of supercells. DFPT is nowadays routinely employed to calculate ground state properties such as force constants and harmonic phonon dispersions \cite{PhysRevB.43.7231,PhysRevLett.58.1861,PhysRevB.51.6773,PhysRevB.55.10355}, dielectric tensors and Born effective charges \cite{PhysRevB.43.7231,RevModPhys.73.515,PhysRevB.55.10355},
the electron-phonon interaction\cite{PhysRevB.54.16487,PhysRevLett.77.1151}, Hubbard parameters\cite{PhysRevB.98.085127}, non-adiabatic phonon dispersions \cite{PhysRevB.82.165111}, just to name a few. DFPT assumes electrons to be in the ground state, occupying the lowest energy Kohn-Sham orbitals and then exploit analytic expressions of the derivative of the total energy with respect to  parameters external to the electronic system such as 
the ionic displacements or applied electric-fields present in the external potential. Its accuracy and 
capabilities are demonstrated by its widespread use in modern condensed matter theory \cite{RevModPhys.73.515}.

The  developments of ultrafast spectroscopic techniques and pump-probe experiments  have opened new perspectives in condensed matter physics and chemistry (see Ref. \onlinecite{doi:10.1021/jacs.9b10533}). In semiconductors, femtosecond laser pulses can promote a substantial number of electrons from valence to conduction; concentrations in excess of 10$^{22}$/cm$^3$ photoexcited carries can be readily achieved\cite{PhysRevB.61.2643}, corresponding to approximately 0.2 excited electrons per atom in silicon. 
As the thermalization of photoexcited carriers typically occurs within hundreds fs, the system experiences an electron-hole plasma at times that are typically smaller of the phonon period (some picoseconds). Until electron-hole recombination takes place (in large gap insulators electron-hole recombination is slower than the typical lattice timescale, the precise value depending on the electron-hole plasma density) the lattice effectively feels a thermalized electron-hole distribution. %(in large gap insulators typical electron-hole recombination times are of the order of the nanosecond)

As the first optically active empty bands of semiconductors are typically composed of antibonding states, 
their occupation determines a substantial variation of the crystal potential and can lead to an ultrafast destabilization of the crystal structure (on the picosecond scale, i.e. much faster of what can be achieved with thermal processes). The sample can then undergo a structural transition to a reversible transient phase that is lost when the laser light is switched off (i.e. the system goes back to its ground state) or it can undergo an irreversible structural transition that remains stable even after that the pulse is removed.
In the first case, the transient phase can be detected either by ultrafast X-ray diffraction at X-ray free electron laser facilities\cite{PhysRevLett.117.135501} or by pump-probe experiments to measure reflectivity, optical absorption, the dielectric function or Raman spectra after the electronic excitation\cite{Ferrante2018,PhysRevLett.80.185}. In the second case, any experimental technique can be used
to characterize the sample as the phase is stable even after removal of the laser source. Examples of this
second class of transition are phase change materials\cite{PhysRevLett.117.135501} and non-thermal melting of semiconductors above critical fluence values\cite{Siders1340,Rousse2001}. A plethora of light-induced phenomena have been observed, including various kinds of electronic phase transitions\cite{mariette2020strain,PhysRevLett.113.026401,PhysRevLett.107.036403}, order-disorder transitions\cite{Wall572}, structural transformations\cite{Fritz633}, light induced charge transfer\cite{Cammarata2021}, detection of warm dense matter\cite{PhysRevLett.106.167601}, phonon softenings\cite{PhysRevLett.108.097401} and non linear phononic effects\cite{Mankowsky2014,Forst2011}. 

This broad range of phenomena calls for efficient theoretical approaches to compute the structural properties of insulators and semiconductors in the presence of a thermalized electron-hole plasma. In this work we develop a constrained DFPT framework (labeled cDFPT) to calculate the forces acting on the ions, the stress tensor (allowing for cell relaxation), the force constant-matrix, the phonon dispersion and the electron-phonon interaction in the presence of a thermalized electron-hole plasma. Our work is based on previous work by  
Tangney and Fahy\cite{PhysRevB.65.054302,PhysRevLett.82.4340}, where the authors developed total energy calculations in the presence of two different chemical potentials, one for the thermalized holes and one for the thermalized electrons.
Moreover, as the presence of a substantial photocarrier concentration (PC) leads to a metallic state, we
generalize the developments of density functional perturbation theory for metals carried out in Ref. \citenum{PhysRevB.51.6773}, to the case of two different chemical potentials for electrons and holes, with particular care for the case of zone center phonons and the treatment of the Fermi shifts induced by the perturbation.

We apply the method to photoexcited tellurium, silicon, gallium arsenide and vanadium diselenide.
We show that, even at the lowest considered PC, the phonon spectrum is substantially
affected and cannot be considered to be the same of the ground state.
We systematically compare our method to the widespread technique of simulating light excitation with a single
Fermi-Dirac distribution with an electronic temperature of the same order of the incoming photon energy (see $e.g.$ Refs.\citenum{PhysRevLett.96.055503} and \citenum{PhysRevLett.77.3149}). We demonstrate that the latter approximation cannot be applied to photoexcited semiconductors and insulators as it leads to inconsistent results when comparing to
experiments, notably when cell relaxation is taken into account. On the contrary our implementation based on two Fermi distributions, one for the thermalized holes and one for the thermalized electrons leads to better agreement with experiments at an affordable computational cost.

The paper is structured as follows. In Sec. \ref{sec2} we illustrate the fundamental assumptions behind the present framework, In Secs. \ref{totens}, \ref{forc}, \ref{vib} we illustrate the mathematical formalism while in Sec.~\ref{app} we present some relevant applications.

\section{Inherent assumptions in the thermalized electron-hole plasma model}\label{sec2}

\begin{figure}[htp]
\centering
\includegraphics[width=1\linewidth]{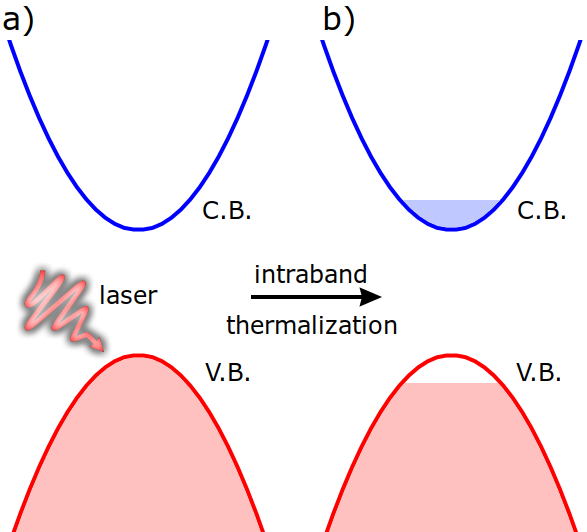}
\caption{Schematic representation of the effect of laser irradiation on electronic occupations. Panel a): electronic distribution before laser irradiation. Panel b): electronic distribution after laser irradiation, after intraband thermalization but before electron-hole recombination takes place. }\label{model}
\end{figure}

The thermalized electron-hole plasma model (also labeled two distribution model in the following) assumes that, for times that are of the order of the phonon period, the photoexcited insulator or semiconductor can be described by a thermalized electron-hole plasma identified by two different chemical potentials (see Fig.~\ref{model}). We are thus assuming that:

(i) photoexcited electron and holes thermalize within hundreds of fs and

(ii) electron-hole recombination times are much longer than the phonon period. \\
These two assumptions are usually satisfied in experiments on insulators and large gap semiconductors as the phonon period is of the order of some picoseconds, while the electron-hole recombination is generally much slower\cite{doi:10.1063/1.108817,doi:10.1063/1.356634}. While the carrier recombination time is known to decrease with increasing density, theoretical arguments and experimental observations indicate that this behaviour saturates at high density for the main recombination mechanisms (radiative and Auger recombination)\cite{doi:10.1063/1.2132524,doi:10.1063/1.1519344,PhysRevB.86.165202}, thus justifying our treatment. This conclusion is also supported by the experiments of Hunsche {\it et al.} on semiconducting tellurium, which demonstrate no significant change in the A$_1$ phonon lifetime as a function of the the carrier density, contrarily to what one would have observed having the electron-hole recombination occurred on a timescale comparable to the phonon period\cite{PhysRevLett.75.1815,PhysRevB.51.6773}.   
In metals, in general, carrier recombination is very fast and the model is inappropriate. However, in some cases, the metal can be seen as a hole-doped insulator and a part of the excitations occurs across an energy gap, thus carrier recombination could still be slow enough for the application of the two distribution model. In this case, however, the validity of the second assumption has to be carefully verified case by case.

We also assume that the excited electron/hole population is large enough to be approximated by an electron-hole plasma. Indeed, while the 
low PC regime is characterized by the occurrence of a resonant peak in correspondence to the exciton energy (see for example~\cite{PL,C5NR00383K,Chernikov2015,doi:10.1142/7184}), when the PC reaches a critical value, the system undergoes a transition towards an electron-hole plasma state~\cite{doi:10.1142/7184,Chernikov2015}
where excitons are not bound anymore and are completely screened. In this regime, the laser pumps electrons
to single-particle conduction-band states.
This requires that the single particle states (i.e. the Kohn-Sham energies) correctly describes the single particle excitations across the gap. It is know that 
DFT with semi-local functionals typically underestimates band gaps (see Tab.\ref{tab1}), and consequently the inclusion of self energy correction like GW may be necessary to obtain a correct description of photoexcitation. However, we argue that the present approach is still valid, as long as the GW correction is well approximated by a scissor operator, implying that the energy correction can be included $a$ $posteriori$ with a rigid shift $\Delta E_{GW}$ of the conduction bands.
 Finally, we underline that in all applications considered in the present paper, we verified that the bottom
 of the conduction band is a bright state. In principle that is not required as the present formalism can also be applied to the case of a dark first excited state ($i.e.$, not allowed by dipole selection rules) by properly initializing the simulation in such a way to fulfill the optical transition matrix element 
 (see Ref.\citenum{PhysRevB.65.054302} for more details). 
 
We now proceed and obtain the expressions for total energy, forces, force constant matrices and electron-phonon coupling in cDFPT.
\begin{table}
    \centering
    \begin{tabular}{|c|c|c|}
    \hline
 Material & gap (th.) & gap (exp.) \\
 \hline
 \hline
    Te & 0.18 & 0.35\cite{Te_gap}\\
    \hline
    Si & 0.45 & 1.17\cite{sc_b}  \\
    \hline
    GaAs & 1.02 & 1.52\cite{sc_b}\\
    \hline
    \end{tabular}
    \caption{Predicted band gaps for bulk tellurium, silicon and GaAs, compared to the experimental value at room temperature. }
    \label{tab1}
\end{table}

\section{Total energy}\label{totens}
 
 The derivation presented here closely follows and extends the one given by De Gironcoli in Ref.\citenum{PhysRevB.51.6773} for the case of metals. We consider the situation where a fraction of the valence electrons e$^{-}_{ph}$ has been excited from the valence to the conduction band. Our hypothesis is that thermal equilibrium is reached separately in both the valence and the conduction band subsets, so that it is reasonable to describe electronic occupations in the two sub-sets with two separate Fermi-Dirac distributions. The smearing function can be a generic smooth approximation to the Dirac's $\delta$ function; in general, different temperatures for valence and conduction distributions are allowed. Similarly to the case of metals, the basic quantity for a practical density-functional theory formulation is the local density of states $n(\mathbf{r},\epsilon)$, which is convoluted with the smearing function $f(\epsilon) = (1/\sigma) \tilde{\delta}(\epsilon/\sigma)$. In the present case, valence and conduction band distributions are separately convoluted with a smearing function. We assume that the first $N_v$ bands are the valence bands. The resulting expression for the local density of states is 
 
\begin{equation}
\begin{gathered}
n(\mathbf{r},\epsilon) = \sum_{i=1}^{N_v} \frac{1}{\sigma} \tilde{\delta}\left(\frac{\epsilon-\epsilon_i}{\sigma}\right)|\phi_i(\mathbf{r})|^2 +  \\
 \sum_{i=N_v+1}^{\infty} \frac{1}{\sigma_c} \tilde{\delta}\left(\frac{\epsilon-\epsilon_i}{\sigma_c}\right)|\phi_i(\mathbf{r})|^2 \enspace ,
 \label{eq1}
\end{gathered}
\end{equation}

where $\epsilon_i$ and $\phi_i(\mathbf{r})$ are the Kohn-Sham eigenvalues and eigenfunctions respectively, $\sigma$ and $\sigma_c$ represent the smearing parameters for the valence and conduction respectively, and the summation over $\mathbf{k}$-points and spin is implicit. We work under the hypothesis that the valence and conduction bands are separated by an energy gap $E_{g}$ much larger than the smearing values, $E_{g} \gg \textrm{max}(\sigma,\sigma_{c})$.  In the case of Fermi-Dirac occupations, $\sigma$ and $\sigma_c$ represent the two temperatures of valence and conduction carriers. The electron density is 

 \begin{equation}
\begin{gathered}
n(\mathbf{r}) = \int_{-\infty}^{\epsilon_F} n(\mathbf{r},\epsilon) d\epsilon + \int_{E_c}^{\epsilon_{F}'} n(\mathbf{r},\epsilon) d\epsilon  \\
 = \sum_i \Theta^{F,F'}_i|\phi_i(\mathbf{r})|^2 \enspace ,
 \label{eq2}
\end{gathered}
\end{equation}

where $E_c$ is the minimum of the conduction, $\epsilon_{F}$, $\epsilon_{F}'$ denote the valence and conduction $quasi$-Fermi levels and the function $\tilde{\Theta}^{F,F'}_{i}$ is case-defined as the integral of the $\tilde{\delta}$ in one of the two band sets:

    \begin{equation}
      \tilde{\Theta}^{F,F'}_{i} = 
    \begin{cases*}
    \tilde{\theta}\left(\dfrac{\epsilon_i-\epsilon_F}{\sigma}\right) & $i=1,N_v$ \\
    \tilde{\theta}'\left(\dfrac{\epsilon_i-\epsilon_{F}'}{\sigma_c}\right) & $i=N_v+1,\infty$ \enspace , \\
    \end{cases*}
     \label{eq3}
\end{equation}

where $\tilde{\theta}(x) = \int_{-\infty}^{x} \tilde{\delta}(y) dy$ and  $\tilde{\theta'}(x) = \int_{E_c}^{x} \tilde{\delta}(y) dy$. The $quasi$-Fermi levels are determined imposing conservation of the number of particles in valence and conduction,  $N_{el}^{v,c}$, separately:
 \begin{equation}
\begin{gathered} N_{el} = N_{el}^v+ N_{el}^c\\
N_{el}^v= \int_{-\infty}^{\epsilon_F } n(\epsilon) d\epsilon = \sum_{i=1}^{N_v} \tilde{\theta}\left(\frac{\epsilon_F-\epsilon_i}{\sigma}\right)\\
N_{el}^c= \int_{E_c}^{\epsilon_F'} n(\epsilon) d\epsilon   = \sum_{i=N_v+1}^{\infty} \tilde{\theta'}\left(\frac{\epsilon_F'-\epsilon_i}{\sigma_c}\right)\\
 \label{eq4}
\end{gathered}
\end{equation}

where n($\epsilon$) is the density of states. In analogy with the case of a metal, the kinetic Kohn-Sham functional $T_s[n]$ is defined through the Legendre transform of the single-particle energy integral, and is written

 \begin{equation}
\begin{gathered}
T_s[n] = \int_{-\infty}^{\epsilon_F}\epsilon n(\epsilon) d\epsilon + \int_{E_c}^{\epsilon_F'}\epsilon n(\epsilon) d\epsilon - \int V_{SCF}(\mathbf{r}) n(\mathbf{r})d\mathbf{r}   \\
= \sum_{i}\left[ -\frac{\hbar^2}{2m} \Theta_i^{F,F'}\mel{\phi_i}{\nabla^2}{\phi_i}+\tilde{\Theta}^{1,F,F'}_i\left(\frac{\epsilon_F-\epsilon_i}{\sigma}\right)\right]  \enspace , \\
 \label{eq5}
\end{gathered}
\end{equation}
\vspace{0.1cm}

where the self consistent potential $V_{SCF}(\mathbf{r})$  is defined in terms of the external potential V($\mathbf{r}$) and the exchange and correlation functional E$_{xc}$ as\cite{RevModPhys.73.515}

\begin{equation}
\begin{gathered}
    V_{SCF}(\mathbf{r}) = V(\mathbf{r})+e^2\int \cfrac{n(\mathbf{r}')}{|\mathbf{r}-\mathbf{r}'|}d\mathbf{r}' + v_{xc}(\mathbf{r})\enspace,\\
    v_{xc} = \dfrac{\delta E_{xc}}{\delta n(\mathbf{r})} 
     \label{eq6}
    \end{gathered}
\end{equation}

and we introduced the function $ \tilde{\Theta}^{1,F,F'}_i$, which is case-defined as

     \begin{equation}
      \tilde{\Theta}^{1,F,F'}_i = 
    \begin{cases*}
    \sigma\tilde{\theta_1}\left(\dfrac{\epsilon_i-\epsilon_F}{\sigma}\right) & $i=1,N_v$ \\
    \sigma_c\tilde{\theta_1}'\left(\dfrac{\epsilon_i-\epsilon_F'}{\sigma_c}\right) & $i=N_v+1,\infty$ \enspace , \\
    \end{cases*}
     \label{eq7}
\end{equation}

where $\tilde{\theta}_1(x) = \int_{-\infty}^{x}y\tilde{\delta}(y)dy$ and 
$\tilde{\theta}'_1(x) = \int_{E_c}^{x}y\tilde{\delta}(y)dy$.\vspace{0.1cm}

The total energy can be written in terms of the kinetic functional

\begin{equation}
    E[n] = T_s[n]+\frac{e^2}{2}\int \frac{n(\mathbf{r})n(\mathbf{r}')}{|\mathbf{r}-\mathbf{r}'|}d\mathbf{r}d\mathbf{r}' + E_{xc}[n]\label{toten}
\end{equation}

and with this choice for the kinetic functional the Kohn-Sham equations follow from the minimization of the total energy with respect to the density, imposing the conservation for the number of valence and conduction electrons. As in the case of the metallic system, the drawback of the smearing approach is that total energy depends on the smearing parameter. In the case of two Fermi levels the error in the total energy is the sum of errors due to the valence and conduction smearing convolution; all the other considerations made for the metallic case \cite{PhysRevB.51.6773} can be extended to this case.

\section{Forces and stress tensor}\label{forc}

With the definition above, forces are computed from the Hellmann-Feynman theorem:

\begin{equation}
    \frac{\partial E}{\partial \mathbf{u}_s} = \int n(\mathbf{r})\frac{\partial V(\mathbf{r})}{\partial \mathbf{u}_s}d\mathbf{r} + \frac{\partial E_{ion}}{\partial \mathbf{u}_s} \enspace .
     \label{eq8}
\end{equation}
where $E_{ion}$ is the electrostatic energy due to the interaction between the ions.

The stress tensor $\alpha_{i,j}$ is defined as minus the derivative of the total energy with respect to to the strain $\eta_{i,j}$ divided by the volume: 

\begin{equation}
    \alpha_{i,j} = -\frac{1}{\Omega}\frac{\partial E_{tot}}{\partial \eta_{i,j}}. \label{eq_stress}
\end{equation}

where the strain is defined as the space scaling operation $\mathbf{r}_i \rightarrow (\delta_{i,j}+\eta_{i,j})\mathbf{r}_{j}$. With the given definition for the kinetic functional, strain calculation is analogous to the metallic case (see Refs. \citenum{PhysRevB.32.3780,PhysRevB.32.3792} for the complete expression of the stress tensor); pressure is defined as minus the trace of the strain tensor, $P = -\sum_i \alpha_{i,i}$.

\section{Vibrational properties}\label{vib}

We now come to the description of lattice dynamics in the two temperature model. The force constant matrix is obtained as the derivative of the Hellmann-Feynman force: 

\begin{equation}
\begin{gathered}
    \mathbf{\Phi}_{s,s'} =   \int \frac{\partial n(\mathbf{r})}{\partial\mathbf{u}_s}\frac{\partial V(\mathbf{r})}{\partial \mathbf{u}_s'} d\mathbf{r} + \int n(\mathbf{r}) \frac{\partial^2  V(\mathbf{r})}{\partial \mathbf{u}_s \partial \mathbf{u}_{s'}} d\mathbf{r}~ +\\
     \frac{\partial^2 E_{ion}}{\partial \mathbf{u}_s \partial \mathbf{u}_{s'}} \enspace .
\end{gathered}
 \label{eq9}
\end{equation}

We then proceed with the calculation of the density variation following an infinitesimal lattice perturbation. We start by considering the generic density response to a perturbation, by direct variation of Eq.(\ref{eq2}):

\begin{equation}
\begin{gathered}
\Delta n (\mathbf{r}) = \sum_i\Theta^{F,F'}_{i}\left[\phi^*_i(\mathbf{r})\Delta\phi_i(\mathbf{r})+c.c.\right] + \\ 
\sum_i|\phi_i(\mathbf{r})|^2\tilde{\Delta}^{F,F'}_{i}\left(\Delta\epsilon^{F,F'}-\Delta\epsilon_i\right) \enspace ,
\end{gathered}
 \label{eq10}
\end{equation}

where  \begin{equation}
      \tilde{\Delta}^{F,F'}_{i} = 
    \begin{cases*}
    (1/\sigma)\tilde{\delta}\left(\dfrac{\epsilon_i-\epsilon_F}{\sigma}\right) & $i=1,N_v$ \\
    (1/\sigma_c)\tilde{\delta}\left(\dfrac{\epsilon_i-\epsilon_F'}{\sigma_c}\right) & $i=N_v+1,\infty$ \enspace , \\
    \end{cases*}
     \label{eq11}
\end{equation}

and  
\begin{equation}
\Delta\epsilon^{F,F'} = 
    \begin{cases*}
    \Delta\epsilon_F & $i=1,N_v$ \\
    \Delta\epsilon_F' & $i=N_v+1,\infty$ \enspace .\\
    \end{cases*}
     \label{eq12}
\end{equation}

We initially neglect the term due to the Fermi shift, $\Delta\epsilon^{F,F'}$  , which will be addressed at the end of the section. The first-order correction to the eigenfunction is orthogonal to the eigenfunction itself and can be expressed in terms of a sum over the spectrum of the perturbed Hamiltonian: 

\begin{equation}
    \Delta\phi_i(\mathbf{r}) = \sum_{j \neq i} \phi_i(\mathbf{r})\dfrac{\mel{\phi_j}{\Delta V_{SCF}}{\phi_i}}{\epsilon_i-\epsilon_j} \enspace ,
     \label{eq13}
\end{equation}

where $\Delta V_{SCF}$ is the variation of the self consistent potential (Eq.(\ref{eq6})) 

\begin{equation}
\Delta V_{SCF}(\mathbf{r}) = \Delta V(\mathbf{r}) + e^2 \int \cfrac{\Delta n(\mathbf{r}')}{|\mathbf{r}-\mathbf{r}'|} d\mathbf{r}' + \cfrac{dv_{xc}(n)}{dn} \Delta n(\mathbf{r}) \enspace .
 \label{eq14}
\end{equation}
\vspace{0.1cm}

We substitute the expression for the wavefunction variation Eq.(\ref{eq13}) in Eq.(\ref{eq10}) and specialize to the case of an infinitesimal lattice perturbation: 

\begin{equation}
\begin{gathered}
\frac{\partial n(\mathbf{r})}{\partial \mathbf{u}_s} = \sum_{i,j} \dfrac{\tilde{\Theta}^{F,F'}_{i}-~\tilde{\Theta}^{F,F'}_{j}}{\epsilon_i-\epsilon_j} \phi_i(\mathbf{r})^*\phi_j(\mathbf{r})\mel**{\phi_j}{\frac{\partial V_{SCF}}{\partial \mathbf{u}_s}}{\phi_i}\enspace.
\end{gathered}
\label{eq15}
 \end{equation}

Using the relation $\tilde{\theta}(x)+\tilde{\theta}(-x) = 1$ and the exchange symmetry between $i$ and $j$ we rewrite 

\begin{equation}
\begin{gathered}
   \frac{ \partial n(\mathbf{r})}{\partial \mathbf{u}_s} = 2 \sum_{i,j}\dfrac{\tilde{\Theta}^{F,F'}_{i}-~\tilde{\Theta}^{F,F'}_{j}}{\epsilon_i-\epsilon_j}\theta_{j,i}\times\\
   \phi_{i}(\mathbf{r})^*\phi_{j}(\mathbf{r})\mel**{\phi_j}{\frac{\partial V_{SCF}}{\partial \mathbf{u}_s}}{\phi_i}
\end{gathered}
\label{eq16}
 \end{equation}
 Here, $\tilde{\theta}_{i,j} = \tilde{\theta}((\epsilon_i -\epsilon_j)/\sigma)$, the first index runs only over the partially occupied states and the second one only over those partially unoccupied. We further simplify the expression by avoiding the sum over the unoccupied states, writing

\begin{equation}
\frac{\partial n(\mathbf{r})}{\partial \mathbf{u}_s}  = 2 \sum_i \phi_i(\mathbf{r})^*\Delta\phi_i(\mathbf{r}) \enspace ,
\label{eq17}
\end{equation}

where $\Delta\phi_i$ satisfy

\begin{equation}
\begin{gathered}
\left[H_{SCF}+Q-\epsilon_i\right]\Delta\phi_i = -\left[\tilde{\Theta}^{F,F'}_{i}-P_i\right]\frac{\partial V_{SCF}}{\partial \mathbf u_s}\ket{\phi_i}\\
 Q = \sum_{k} \alpha_{k}\ket{\phi_k}\bra{\phi_k}\\
 P = \sum_j\beta_{i,j}\ket{\phi_j}\bra{\phi_j}\\
 \beta_{i,j} = \tilde{\Theta}^{F,F'}_{i}\tilde{\theta}_{i,j}+\tilde{\Theta^{F,F'}_i}\tilde{\theta}_{i,j}+\alpha_j\dfrac{\tilde{\Theta}_{F,i}-~\tilde{\Theta}_{F,j}}{\epsilon_i-\epsilon_j}\tilde{\theta}_{j,i} \enspace .
\end{gathered}
\label{eq18}
 \end{equation}

where H$_{SCF}$ is the unperturbed Kohn-Sham Hamiltonian\cite{RevModPhys.73.515}. Here, $\alpha_k$'s are chosen in such a way that the~$Q$ operator makes the linear system Eq.(\ref{eq18}) non-singular for all non-vanishing $\Delta\phi_k$. A possible simple choice is $\alpha_k= \mathrm{max}(\epsilon_F'+\Delta-\epsilon_k,0)$. Another choice is to set $\alpha_k$ equal to the occupied bandwidth plus a certain quantity, $e.g.~3\sigma$, for all (partially) occupied states and equal to zero for totally unoccupied states. 

\subsection{Zone center phonons inducing Fermi shifts}

In this section we discuss the peculiarities related to the
calculation of the linear response for optical
zone-center phonons.
In DFPT (and cDFPT) codes
the phonon calculation with ${\bf q}\ne {\bf 0}$  and
${\bf q}={\bf 0}$ are
treated with two different approaches.
At ${\bf q}\ne {\bf 0}$, the calculation is performed within the grand canonical
ensemble, with a (two in our case) constant(s) electron chemical potential(s).
At ${\bf q}={\bf 0}$, the calculation is performed in the canonical ensemble
with a constant number of electrons. When the phonon displacement induces a change in the number of the electrons a Fermi shift has to be included to enforce charge conservation. In our framework this amounts at introducing two Fermi shifts for the 
 two $quasi$-Fermi levels in the density response:
 
 \begin{equation}
\begin{gathered}
\frac{\partial n(\mathbf{r})}{\partial \mathbf{u}_s}  = 2 \sum_i \phi_i(\mathbf{r})^*\Delta\phi_i(\mathbf{r})  + \\
\sum_i|\phi_i(\mathbf{r})|^2\tilde{\Delta}^{F,F'}_{i}\left(\Delta\epsilon^{F,F'}-\Delta\epsilon_i\right) \enspace .
\end{gathered}
\label{eq19}
 \end{equation}

 This term can only arise for a periodic perturbation at $\mathbf{q} = 0$, in compounds where the atomic positions are not all fixed by symmetry.
 
 The Fourier transform of the self consistent potential variation, $\Delta V_{SCF} (\mathbf{q})$, reads 
 \begin{equation}
  \Delta V_{SCF}(\mathbf{q}) = \frac{4\pi e^2}{q^2}\Delta n_{ext}(\mathbf{q}) + \frac{4\pi e^2}{q^2}\Delta n(\mathbf{q}) + \frac{dv_{xc}}{dn}\Delta n(\mathbf{q}) \enspace ,
  \label{eq20}
 \end{equation}
  \vspace{0.1cm}
  
  which we separate into the contributions due the valence and conduction electrons  $\Delta V^{v,c}_{SCF}(\mathbf{q})$:
   
   \begin{equation}
 \begin{gathered}
  \Delta V_{SCF}(\mathbf{q}) = \Delta V^v_{SCF}(\mathbf{q}) + \Delta V^c_{SCF}(\mathbf{q})\\
 \Delta V^v_{SCF}(\mathbf{q}) = \frac{4\pi e^2}{q^2}\Delta n^v_{ext}(\mathbf{q}) + \frac{4\pi e^2}{q^2}\Delta n^v(\mathbf{q}) + \frac{dv_{xc}}{dn}\Delta n^v(\mathbf{q})\\
 \Delta V^c_{SCF}(\mathbf{q}) = \frac{4\pi e^2}{q^2}\Delta n^c_{ext}(\mathbf{q}) + \frac{4\pi e^2}{q^2}\Delta n^c(\mathbf{q}) + \frac{dv_{xc}}{dn}\Delta n^c(\mathbf{q}) \enspace ,
 \label{eq21}
 \end{gathered}
   \end{equation}
   \vspace{0.1cm}

where $\Delta n_{ext}^{v,c}(\mathbf{q})$ represents the valence and conduction density variation due to the macroscopic electrostatic component of the perturbing potential. Similarly, we decompose the density variation induced by a generic perturbation, $\Delta n(\mathbf{q})$, into the valence and conduction terms  $\Delta n^{v,c}(\mathbf{q})$ :

\begin{equation}
 \begin{gathered}
 \Delta n(\mathbf{q}) = \Delta n^v(\mathbf{q}) + \Delta n^c(\mathbf{q}) \\
 \Delta n^v(\mathbf{q}) = -n(\epsilon_F)\Delta V^{v}_{SCF}(\mathbf{q}) + \Delta n^{lf}_v (\mathbf{q})\\
 \Delta n^c(\mathbf{q}) = -n(\epsilon_F')\Delta V^{c}_{SCF}(\mathbf{q}) + \Delta n^{lf}_c (\mathbf{q}) \enspace . \\
\end{gathered}
\label{eq22}
\end{equation}

where $\Delta n^{lf}_{v,c}$ is the valence(conduction) part of the density response to the non macroscopic component of the self consistent potential. The self consistent potential variation due the valence and conduction electrons can then be expressed as:
\begin{equation}
\begin{gathered}
    \Delta V^{v}_{SCF}(\mathbf{q}) = - \dfrac{ \Delta n_{v}(\mathbf{q}) - \Delta n^{lf}_v(\mathbf{q})}{n(\epsilon_F)}\\
    \Delta V^{c}_{SCF}(\mathbf{q}) = - \dfrac{ \Delta n_{c}(\mathbf{q}) - \Delta n^{lf}_c(\mathbf{q})}{n(\epsilon_{F}')} \enspace .
\end{gathered}
\label{eq23}
\end{equation}
 For $\mathbf{q}\approx 0$, we have $\Delta n_{ext}^{v,c}(\mathbf{q}) = \Delta n^{v,c}(\mathbf{q}) + \mathcal O(q^2)$. We enforce charge neutrality applying $quasi$-Fermi levels shifts equal and opposite to the variation of the macroscopic component of the self consistent potential felt by valence and conduction electrons:

\begin{equation}
\begin{gathered}
\Delta \epsilon_F = \dfrac{\Delta n_{ext}^v(\mathbf{q} = 0) -\Delta n^{lf}_v(\mathbf{q=0})}{n(\epsilon_F)}\\
\Delta \epsilon_{F}' = \dfrac{\Delta n_{ext}^c(\mathbf{q} = 0) - \Delta n^{lf}_c(\mathbf{q=0})}{n(\epsilon_F')} \enspace .
\end{gathered}
\label{eq24}
\end{equation}

 For a neutral external perturbation like an infinitesimal lattice perturbation, atoms are displaced but the total charge does not change, $i.e.$ $\Delta n_{ext}^{v,c} = 0 $. We thus obtain the final expression for the $quasi$-Fermi levels shifts

\begin{equation}
\begin{gathered}
\Delta \epsilon_F = - \dfrac{\Delta n^{lf}_v(\mathbf{q=0})}{n(\epsilon_F)}  = \cfrac{\int n(\epsilon_F,\mathbf{r})\cfrac{\partial V_{SCF}(\mathbf{r})}{\partial \mathbf{u}_s} d\mathbf{r}}{n(\epsilon_F)} \\
\Delta \epsilon_{F}' = - \dfrac{\Delta n^{lf}_c(\mathbf{q=0})}{n(\epsilon_F')} = \cfrac{\int n(\epsilon_F',\mathbf{r})\cfrac{\partial V_{SCF}(\mathbf{r})}{\partial \mathbf{u}_s} d\mathbf{r}}{n(\epsilon_F')} \enspace .
\end{gathered}
\label{eq25}
\end{equation}

We rewrite Eq.(\ref{eq10}) as

\begin{equation}
\begin{gathered}
\Delta'\phi_i(\mathbf{r}) = \Delta\phi_i(\mathbf{r}) + \Delta^{sh}\phi_i(\mathbf{r}) \enspace ,
\end{gathered}
\label{eq26}
\end{equation}
\vspace{0.1cm}

where we defined the wavefunction shift as 

\begin{equation}
    \Delta^{sh}\phi_i(\mathbf{r}) = (1/2) \Delta\epsilon^{F,F'} \phi_i(\mathbf{r}) \enspace .
    \label{eq27}
\end{equation} 

The compact expression for the electron density perturbation is then:

\begin{equation}
\begin{gathered}
\frac{\partial n(\mathbf{r})}{\partial \mathbf{u}_s}  = 2 \sum_i \phi_i(\mathbf{r})^*\Delta'\phi_i(\mathbf{r})
\end{gathered}
\label{eq28}
\end{equation}

\subsection{Electron-phonon interaction}

The last part of our derivation is concerned with the extension of the electron-phonon matrix elements calculation to the case of the two distribution model. The deformation potential corresponding to a lattice perturbation of momentum $\mathbf{q}$ and irreducible representation $\nu$ is written

\begin{equation}
d^{m,n}_{\mathbf{k},\mathbf{q},\nu} = \bra{\phi_{\mathbf{k},m}}\frac{\partial V_{SCF}}{\partial u_{\mathbf{q},\nu}}\ket{\phi_{\mathbf{k}+\mathbf{q},n}} \label{eq31}
\end{equation}

where $\ket{\phi_{\mathbf{k},m}}$ represents a Kohn-Sham eigenvector characterized by momentum $\mathbf{k}$ and band index $m$. In the case of two separate carrier distribution, the procedure remains essentially the same, exception made for the case $\mathbf{q}=\mathbf{0}$, and the presence of a non-vanishing Fermi-shift for either valence or conduction (or both). In this case, the  $\mathbf{q}=\mathbf{0}$ component of the calculated potential variation has to be shifted according to Eq.\ref{eq23} in order to maintain charge neutrality separately in valence and conduction. Once this quantity is calculated, the electron-phonon matrix elements follow as in standard DFPT\cite{RevModPhys.73.515}.

\section{Applications}\label{app}

\subsection{Tellurium}

We apply the presented formalism to the case of bulk tellurium. At ambient conditions the stable phase is $\alpha$-Te, consisting of helical chains parallel to the c axis of the trigonal $P3_{1}21-D_{3}^{4}$ structure\cite{PhysRevB.65.054302}. 
%. 
Bulk tellurium has been previously studied in the framework of two carriers distribution model in Ref.\citenum{PhysRevB.65.054302}. This study attributed the experimentally observed reflectivity oscillations under photoexcitation to the so called displacive excitations of coherent phonon (DECP) mechanism, caused by the A$_1$ phonon mode. The motion along the A$_1$ phonon eigenvector corresponds to a modulation of the free internal coordinate $x$, corresponding to the helical chain radius. Since the  A$_1$ mode involves the motion of a free internal coordinate, a Fermi shift must occur \cite{PhysRevB.82.165111}. 
As such, $\alpha$-Te gives us the chance to verify the consistency of our implementation, comparing phonon eigenvalues obtained in finite differences calculations and the one obtained in cDFPT.

We proceed in the following way: first, we calculate the ground state structure by performing a variable-cell relaxation structural parameters (see Tab.\ref{tab2} for the structural parameters). We then explore two possible cases: in a first scenario we keep the volume of the unit cell fixed under photoexcitation, as was previously done in Ref.\citenum{PhysRevB.65.054302}; in a second scenario, we also include the volume change induced by pumping. The reason is the following: the lattice relaxation due to electron photoexcitation can be thought as an instantaneous change of the Born-Oppenheimer potential felt by the nuclei, consequently the lattice readjustment is a purely electronic effect. As such, we expect the crystal to readjust to the new lattice parameters in a short timescale, much shorter than the thermal one, as demonstrated by multiple experimental observations\cite{cryst11020186,Siders1340,Rousse2001}. While the precise timescale for non-thermal lattice readjustment depends on size and composition of the sample, from the experimental observations we infer that the typical timescale for a 100 nm sized silicon sample is less than 1 ps\cite{cryst11020186}, thus it is physically meaningful to perform variable-cell relaxation at the considered PC value and perform the phonon calculation on the volume-relaxed cell.

\begin{figure}[htp]
\centering
\includegraphics[width=1\linewidth]{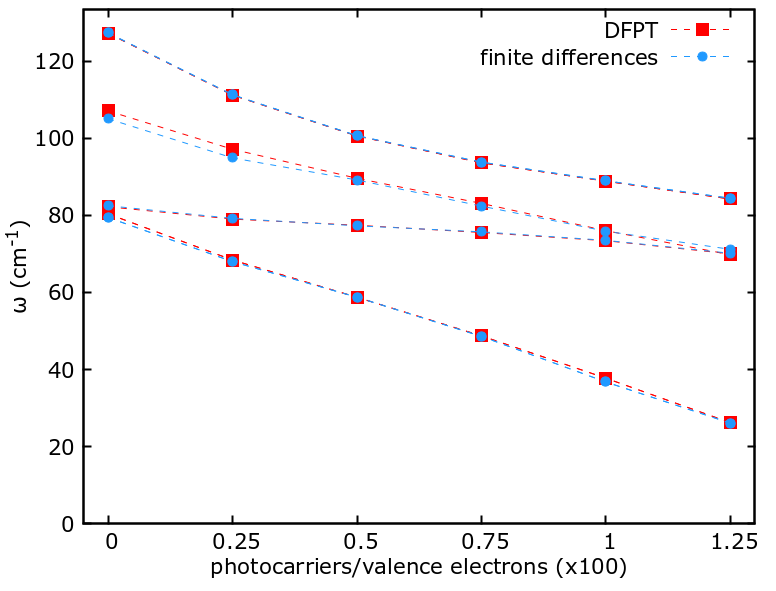}
\caption{Phonon frequencies for bulk tellurium at $\Gamma$-point as a function of PC, without the effect of volume relaxation. Red lines indicate the results obtained within cDFPT, while blue lines are calculated with the finite differences method.}\label{tefig1}
\end{figure}

\begin{figure}[htp]
\centering
\includegraphics[width=1\linewidth]{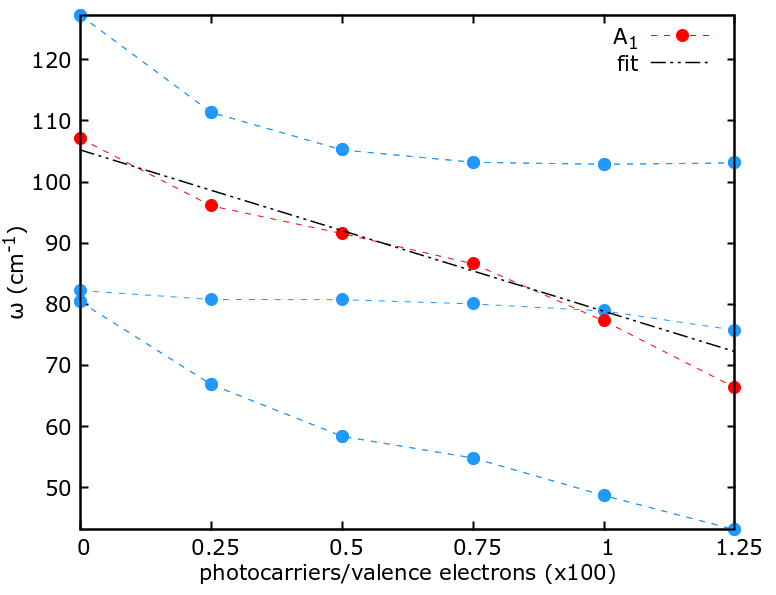}
\caption{Phonon frequencies for bulk tellurium at the $\Gamma$-point as a function of PC, including the effect of volume relaxation. Red line indicate the A$_1$ symmetry phonon responsible for DECP, the black line represents the corresponding linear fit.}\label{tefig2}
\end{figure}

\begin{table}
    \centering
    \begin{tabular}{|l|c|c|c|c|c|}
    \hline
Parameter & \multicolumn{5}{c|}{Photocarriers per Te atom}\\ \hline
    & 0\% & 0.25\% &  0.5\% & 0.75 \% &  1\%\\\hline
 \hline
  & & & & & \\[-1em] 

a(Te) (\AA) & 4.502 & 4.506 & 4.52 & 4.533 & 4.546  \\\hline
  & & & & & \\[-1em] 
c/a(Te) & 1.325 & 1.31 & 1.29 & 1.27 & 1.25 \\\hline
  & & & & & \\[-1em] 
x(Te) (crystal) & 0.271 & 0.275 & 0.281 & 0.287 & 0.293 \\\hline
\hline
  & & & & & \\[-1em] 
  & 0 e$_{ph}^-$ &  0.1 e$_{ph}^-$&  0.2 e$_{ph}^-$ & 0.3 e$_{ph}^-$&  0.4 e$_{ph}^-$\\\hline
  & & & & & \\[-1em] 
  a(Si)(\AA)  & 5.381 & 5.372 & 5.365 & 5.361 & 5.359 \\\hline
   & & & & &  \\[-1em] 
 a(GaAs) (\AA) & 5.527 & 5.541 & 5.561 & 5.583 & 5.623 \\\hline
    \end{tabular}
    \caption{Structural parameters as a function of PC for silicon, gallium arsenide and tellurium in the cDFPT approach. }
    \label{tab2}
\end{table}

\begin{figure*}[htp]
\centering
\includegraphics[width=0.8\linewidth]{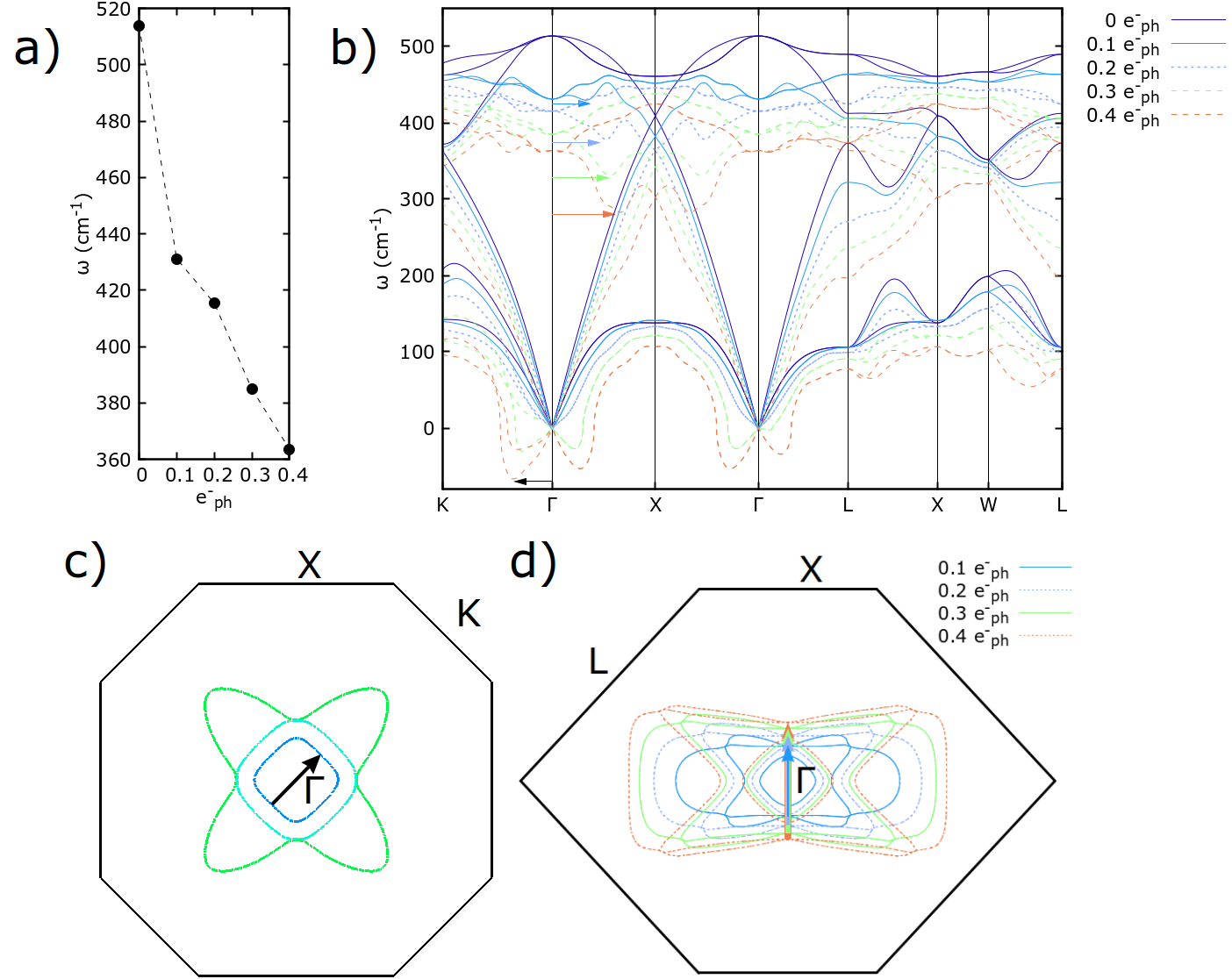}
\caption{Panel a): optical mode frequency at $\Gamma$ as a function of PC. Panel b): phonon frequencies for silicon at selected PC values. Panel c): section of the valence FS at 0.4 e$^-_{ph}$/z.c. concentration, in the plane perpendicular to the [110] reciprocal direction. The black arrow indicates the nesting vector responsible for the phonon instability. Panel d: section of the valence FS at various PC values, in the plane perpendicular to the [211] reciprocal direction. Arrows indicate nesting vectors responsible for the softenings observed in the optical branch along the $\Gamma$-X direction. FS plots have been realized using the open source software FermiSurfer\cite{KAWAMURA2019197}. }\label{sifig3}
\end{figure*}

First of all, we use the case of tellurium as a benchmark for our implementation, comparing the phonon frequencies obtained within cDFPT with the ones obtained with a finite differences calculation\cite{PHON}. The comparison is reported in Fig.~\ref{tefig1} demonstrating an excellent agreement between the results in the two approaches, confirming the consistency of our implementation. 

Without including volume effects, we obtain a phonon frequency shift of -0.91 THz per 1\% excitation of the valence population. Using the model in Ref.\citenum{PhysRevB.65.054302} and assuming linearity between the reflectivity changes and phonon frequency shift, we infer that the derivative of the reflectivity peak frequency with respect to pump fluence is equal to -0.08 THz per mJ/cm$^2$, against the experimental value of -0.07 THz per mJ/cm$^2$ reported in Ref.\citenum{PhysRevB.65.054302}, obtained by fitting the low excitation part of the experimental data presented in Ref.\citenum{PhysRevLett.75.1815}. Thus, neglecting cell relaxation, the value is underestimated by $\approx 14\%$.

We then proceed to include the effect of the volume change in the frequency calculation. We relax the tellurium structural parameters at each PC. The structural parameters as a function of PC are reported in Tab.~\ref{tab2}, while phonon frequencies behavior at $\Gamma$ is shown in Fig.~\ref{tefig2}. We fit the A$_1$ mode in the low photoexcitation regime, between 0\% and 0.75\%  carrier concentration, obtaining a phonon frequency shift of -0.79 THz per 1\% excitation of the valence population. In the same assumptions as before, we obtain that the derivative of the reflectivity peak frequency with respect to pump fluence is -0.0693 THz per mJ/cm$^2$, in excellent agreement with the experimental value of 0.07 THz per mJ/cm$^2$ (an overestimation of $1\%$). This result implies that the inclusion of cell relaxation effects must be included in order to obtain a quantitative agreement with the experimental observations

\subsection{Silicon}

\begin{figure}[htp!]
\centering
\includegraphics[width=1\linewidth]{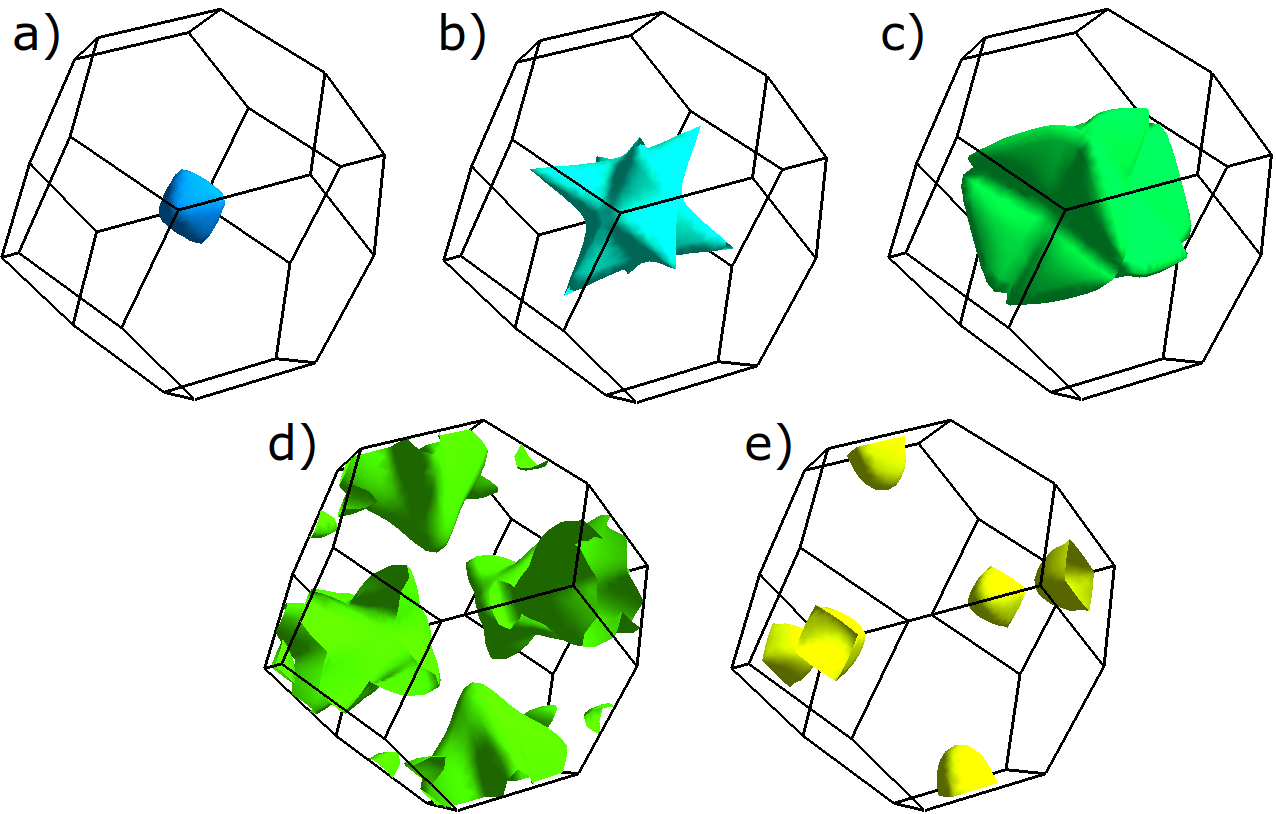}
\caption{Panel a), panel b) and panel c): the three sheets of the valence $quasi$-FS for silicon at a PC value of 0.4 e$^-_{ph}$/z.c.. Panel d) and panel e): the two sheets of the conduction $quasi$-FS at a PC value of 0.4 e$^-_{ph}$/z.c.. FS plots have been realized using the open source software FermiSurfer\cite{KAWAMURA2019197}.}\label{sifs}
\end{figure}

We switch to the study of the vibrational response of silicon under photoexcitation in the cDFPT approach. According to the discussion made for tellurium, calculations are performed on the volume-relaxed cell at the investigated PC value. We start discussing the phonon frequencies behavior as a function of the PC, in panels a) and b) of Fig.~\ref{sifig3}. The behavior of the topmost phonon frequency at the $\Gamma$-point as a function of PC is reported in panel a) of Fig.~\ref{sifig3}. 
We observe a $\approx 16.1\%$ softening of optical phonon frequencies already at a PC value of 0.1 e$^-$ per zincblende cell (z.c.), demonstrating that at values of PC large enough to screen excitonic effects but substantially lower than the largest PCs that are achieved in experiments,  the vibrational response of the system is substantially affected and the phonon spectrum cannot be considered as frozen (i.e. unchanged with respect to the unexcited case). In panel b) of Fig.~\ref{sifig3}, we report the full phonon spectrum for silicon along high symmetry directions. We point out that no acoustic sum rule was enforced in the calculation of phonon eigenvalues. 
Softenings can be observed near the $\Gamma$ point with increasing PC, signaling the progressive formation of a structural instability. We recall that close to the structural instability, anharmonic effects could become relevant. 

We conclude pointing out that while conventional melting occurs heterogeneously at high atomic mobility (thermodynamic melting), experimental evidences show that an homogeneous mechanical melting under irradiation can be observed, and that a competition between thermodynamic and mechanical melting exists\cite{M1,Wolf1990}. The development of imaginary phonon frequencies is related to a second order phase transition and thus to mechanical melting, however, it is in general possible that a {\it latent} first order transition occurs before the phonon softening.

\begin{figure}[htp!]
\centering
\includegraphics[width=1\linewidth]{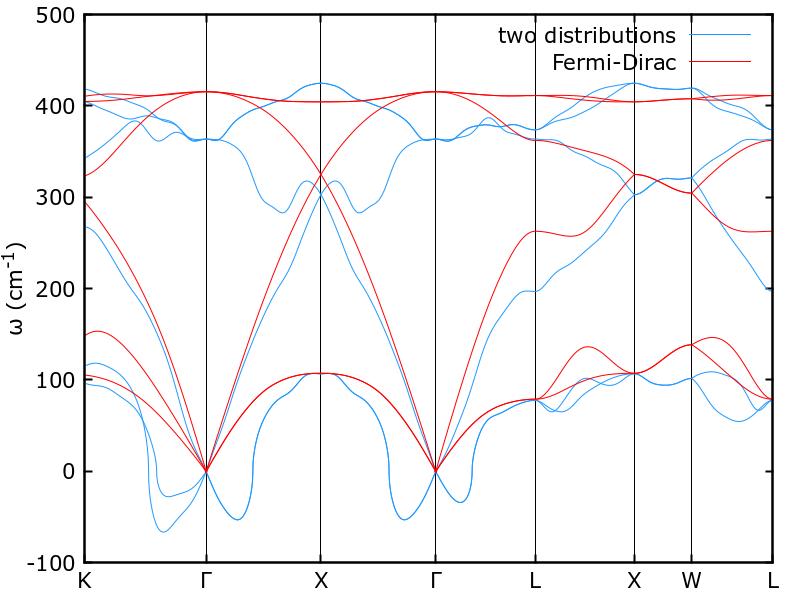}
\caption{Phonon frequencies for silicon at 0.4 e$^-$/z.c. obtained with the two carrier distribution approach and with the single Fermi-Dirac distribution. In the case of the two carrier distribution approach, we observe the appearance of softenings, while no softenings can be seen in the case of a single Fermi-Dirac distribution approach.}\label{sifig1}
\end{figure}

\begin{figure*}[htp]
\centering
\includegraphics[width=0.8\linewidth]{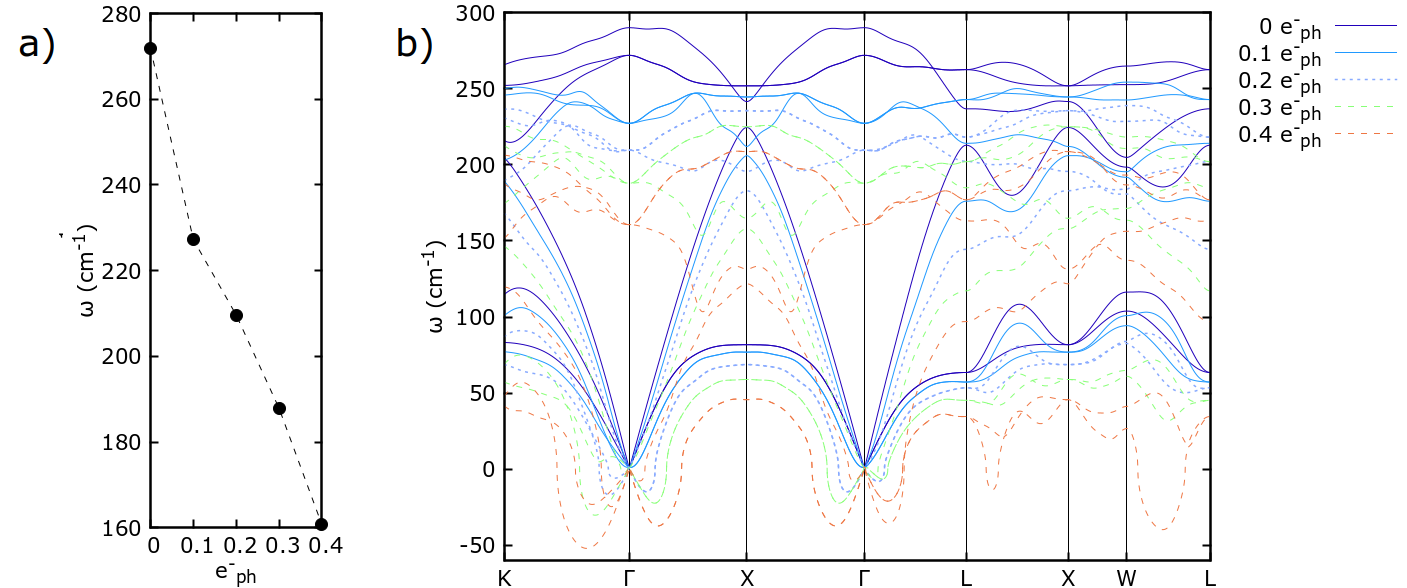}
\caption{Panel a): optical phonon frequency at $\Gamma$ at increasing PC values. Panel b):  phonon frequencies for GaAs for selected PC values. Phonon softenings form near $\Gamma$ for PC values up to 0.3 e$_{ph}^-$/$f.u.$, while new softenings along the $L-X$ and $W-L$ high symmetry directions emerge at 0.4 e$_{ph}^-$/$f.u.$.}\label{gaasfig1}
\end{figure*}

At any finite PC values, we observe the formation of multiple softenings in the phonon branches, which here we address in more details. The full band-resolved valence and conduction Fermi surfaces (that we label {\it $quasi$-FSs}) at a PC value of 0.4 e$^-_{ph}$/z.c. are reported in Fig.~\ref{sifs}. We show that the valence $quasi$-FS is mainly responsible for phonon softenings. This is clearly illustrated in panels c) of Fig.~\ref{sifig3} where we show a cut of the valence $quasi$-FS perpendicular to the [110] reciprocal direction. We relate the vector nesting the smaller section of the valence $quasi$-FS along the $\Gamma-K$ direction to the main phonon softening observed in the phonon spectrum at 0.4 (see panels b) and c) of Fig.~\ref{sifig3}, black arrow). We further relate optical branch softenings to a specific nesting vector for the valence $quasi$-FS sheets depicted in Fig.\ref{sifs} b) c). We perform cuts of the valence $quasi$-FS perpendicular to the [211] reciprocal direction as a function of PC, finding that the optical branch softening in the $\Gamma-X$ direction happens at the wave-vector where the two external sheets of the valence $quasi$-FS touch (see panel b) and d) of Fig.~\ref{sifig3}, colored arrows).

Previous theoretical studies\cite{PhysRevLett.77.3149,PhysRevLett.96.055503} reported that silicon under intense photoexcitation is subject to lattice instabilities and structural changes, in agreement with the experimentally observed melting of silicon\cite{PhysRevLett.51.900}. However, the transition mechanism is not completely clear. In the $ab$-$initio$ molecular dynamical study by Silvestrelli $et$ $al.$ \cite{PhysRevLett.77.3149} it was shown that a high electronic temperature of 2.15 eV causes the melting of the crystal. It was claimed that no soft phonon frequencies were observed. In the DFPT study by Recoules $et$ $al.$\cite{PhysRevLett.96.055503}, considering the same electronic temperature of 2.15 eV, soft phonon modes were observed along most of the high symmetry direction of the zincblende crystal, in contradiction with Ref. \onlinecite{PhysRevLett.77.3149}. In both studies, volume was kept fixed to its equilibrium value. However, there is no real reason for this assumption as, after the pumping and after the consequent thermalization of the electronic system, when the lattice degrees of freedom sets in, the cell deformation does occur \cite{cryst11020186}. 

In order to better clarify this issue, 
in Fig.~\ref{sifig1} we compare the calculated phonon spectrum for silicon at 0.4 e$^-_{ph}$/z.c. in cDFPT (blue lines) with the one obtained within a  high temperature Fermi-Dirac approach (red lines), employing an electronic temperature of 1 eV, corresponding to 0.4 e$^-_{ph}$/z.c. photocarriers. In both cases, the lattice parameter is determined with a variable-cell relaxation procedure. As expected, the single Fermi-Dirac distribution approach leads to smooth phonon dispersion curves with all the quasi-FS softenings washed out by the large Fermi temperature. Furthermore, no unstable phonons occurs.
On the contrary, in the two Fermi distributions approach, as the holes and electrons temperatures are low, the phonon dispersion still display quasi-FS related softenings and the occurrence of a structural instability.
Moreover, the softening of the optical mode at zone center is substantially larger in the case of Si. 
On the contrary, we observe a $\approx 1 \%$ volume reduction in the cDFPT approach at 0.4 e$^-_{ph}$/z.c., in agreement with the experimental observation of Ref. \onlinecite{cryst11020186}, while a 3\% volume increase is predicted by the high temperature Fermi-Dirac approach.
The differences in volume expansion with the two approaches are even more severe at higher fluences.
Indeed, at the fluences of 2.15 eV considered in Refs. \onlinecite{PhysRevLett.77.3149,PhysRevLett.96.055503}
the use of an electronic temperature of 2.15 eV leads to a lattice expansion exceeding 10\% of the original lattice parameter ( $5.98$ \AA~ against the low temperature LDA value of $5.38$ \AA ). On the contrary,
the use of a converged electronic temperature in the framework of the two distribution model leads to a 
$2\%$ lattice expansion ($5.48$ \AA).

\begin{figure}[htp]
\centering
\includegraphics[width=1\linewidth]{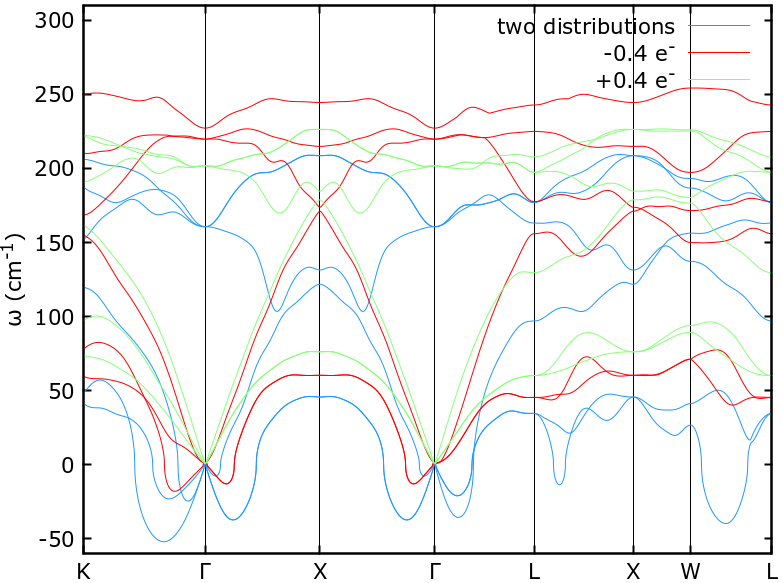}
\caption{Phonon frequencies for GaAs with the two carrier distribution at 0.4 e$^-_{ph}$/$f.u.$ (blue lines), compared with phonon frequencies obtained with a doping of +0.4 e$^-$ (green lines) and  -0.4 e$^-$ (blue lines).}\label{gaasfig4}
\end{figure}

On a general basis, we can thus conclude, that the use of a single Fermi Dirac distribution with temperatures of the order of the incoming laser frequency, leads to unrealistic values of the volume expansion. In this framework, the approximation of considering the cell fixed in Si, beside being unjustified (at least before the thermal melting occurs), artificially enhances the tendency towards phonon instabilities with respect to the case in which, in the same framework, the structural optimization is carried out.  

Finally, the complete structural optimization using two distributions each one with cold carriers leads to stronger phonon softening and preserves the Kohn-anomalies due to Fermi surface nesting of conduction of valence quasi Fermi Surfaces. Moreover it generally gives a better agreement with experiments with respect to the case of a single Fermi-Dirac distribution with hot electrons.

\subsection{Gallium arsenide}

\begin{figure*}[htp]
\centering
\includegraphics[width=0.8\linewidth]{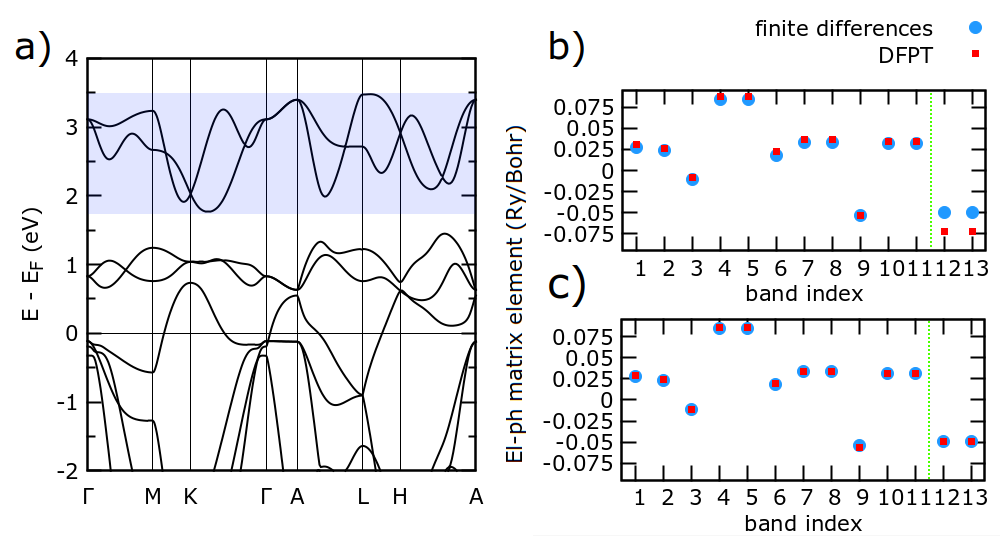}
\caption{Panel a): band structure for bulk VSe$_2$. The semi-transparent blue rectangle highlights the conduction states considered in the cDFT and cDFPT simulations. Panel b) and c): diagonal deformation potential matrix elements $d^{n,n}_{\mathbf{k},\mathbf{q=0}}$ along the out-of-plane coordinate of one selenium atom, with 0.25 e$_{ph}^-$/$f.u.$ moved to the conduction states, without (b) and with (c) Fermi shift. The green dashed lines separate valence and conduction regions of the cDFT and cDFPT simulation.}\label{vse2fig1}
\end{figure*}

We now consider the vibrational response of GaAs under photoexcitations in cDFPT. Experimental observations demonstrate that crystalline GaAs under irradiation undergoes an insulator-metal transition for high fluence values, while also showing a structural amorphization of non-thermal origin\cite{PhysRevB.66.245203}. We calculate the phonon spectrum at varying PC, exploring the range 0-0.4 e$^-_{ph}$/$f.u.$, corresponding to 0-10\% valence electron excitation. Results are depicted in Fig.~\ref{gaasfig1}. In panel a) the behavior of $\Gamma$ acoustic phonon frequency is reported (no LO-TO splitting is included here), demonstrating a sizable effect also a low PC. As it was observed for Si, even at the lowest PC of 0.1 e$^-_{ph}$/$f.u.$, the phonon spectrum is substantially affected and cannot be considered unchanged with respect to the undoped case. 

In the phonon dispersion reported in panel b),  we also include LO-TO splitting at $\Gamma$ in absence of photocarriers, while we don't include it at finite PC values, assuming that the screening effect of photoexcited carriers will suppress it at any finite PC value. Similarly to what observed for silicon,  phonon softenings emerge close to the $\Gamma$ point of the BZ for increasing PC values, pointing towards the formation of lattice instabilities. Other softenings are observed in the $W-L$ reciprocal direction and the $L-X$ reciprocal direction. As in silicon, anharmonic effects may change the critical fluence for the formation of a phonon instability. Interestingly, we observe that volume increases as the PC is increased in the range 0-0.4 e$^-$/$f.u.$, at variance with what we observe in silicon. We point out that no acoustic sum rule was enforced in the calculation of phonon eigenvalues.  

Differently from what we have done for silicon, here we disentangle the role of the two $quasi$-FSs by comparing the calculated cDFPT phonon spectrum with the ones obtained by a negative (positive) doping of 0.4 e$^-$ (-0.4 e$^-$) performed on the same crystal structure. Results are shown in Fig.~\ref{gaasfig4}. We observe that similar softenings around the $\Gamma$ point, although less pronounced, appear for a positive doping of -0.4e$^-$, thus we conclude that the observed softenings are caused by the valence $quasi$-FS also in GaAs. The close correspondence to the results shown for silicon is not surprising as the electronic structure of GaAs (not shown) closely resembles the one of silicon, in particular its valence $quasi$-FS contains the same $quasi$-FS shown in Fig.~\ref{sifs} a), which is responsible for the phonon instability. 

\subsection{Vanadium diselenide}

Finally, we demonstrate the reliability of electron-phonon coupling matrix element calculations by considering the case of the bulk VSe$_2$ in its stable polytype (1T). VSe$_2$ is has metallic character, nevertheless the cDFPT formalism can still be applied, provided that the photocarrier population in conduction is separated from the lower states by an energy gap; this aspect is highlighted in the band structure reported in Fig.~\ref{vse2fig1} a). We choose VSe$_2$ since it possesses a free internal coordinate in the out-of plane selenium position, and thus a non-zero Fermi shift (or better, two non-zero Fermi shifts in the case of photoexcitations). Furthermore, this material has sizable electron-phonon matrix elements. We exploit the following relation existing between the deformation potential $d^{n,m}_{\mathbf{k},\mathbf{q}}$ and Kohn-Sham eigenvalues 

\begin{equation}
d^{n,n}_{\mathbf{k},\mathbf{q=0},\nu} = \frac{d\epsilon_{\mathbf{k},n}}{du_{\mathbf{q=0},\nu}}
\end{equation}

where n is the band index, while $du$ represents the infinitesimal motion of atoms along a generic pattern in the primitive cell. We calculate the deformation potential elements in cDFPT and the eigenvalue derivative in the finite differences approach. Our results are reported in Fig.~\ref{vse2fig1} in panel b) and c), for 0.25 e$_{ph}^-/f.u$. excitation, without and with the Fermi shift correction, respectively. The error committed without including the Fermi shift is especially evident for the conduction electron-phonon matrix elements. We verified that the error committed in panel b) corresponds exactly to the shift of the valence/conduction $quasi$-Fermi levels induced by the perturbation. The excellent agreement between the finite differences method and the cDFPT calculation demonstrated in panel c) confirms the consistency of our formulation. 

\section{Methods}
\label{Methods}

First-principles calculations have been performed within density functional theory. We have used scalar-relativistic Optimized Norm-Conserving Vanderbilt pseudopotentials\cite{doi:10.1021/acs.jctc.6b00114} to describe the electron-ion interaction in the case of tellurium and vanadium diselenide, and Hartwigsen-Goedecker-Hutter\cite{PhysRevB.58.3641} in the case of gallium arsenide and silicon, while employing a kinetic energy cutoff of 60~Ry (80~Ry for VSe$_2$) in the plane-wave expansion of the Kohn-Sham wavefunctions to converge the stress tensor in the variable-cell calculations, as implemented in the \textit{Quantum Espresso} (QE) package\cite{QE,QE2}. The presence of the two $quasi$-Fermi surfaces was dealt with using the smearing approach. Two Fermi-Dirac distributions with smearing parameters $\sigma$=$\sigma_c$= 0.05 eV were considered for Te, while the Marzari-Vanderbilt approach\cite{PhysRevLett.82.3296} was employed for Si,GaAs and VSe$_2$, with valence and conduction smearing parameters $\sigma$=$\sigma_c$=0.136 eV.  For the exchange-correlation potential we have adopted the generalized gradient approximation (GGA) in the Perdew, Burke and Ernzerhof (PBE)\cite{PBE} formulation in tellurium and vanadium diselenide, while local density approximation (LDA)\cite{PZ}  was adopted in silicon and gallium arsenide. In order to ensure  a proper relaxation procedure, forces have been relaxed down to $1\times10^{-5}$ Ry/Bohr. A {16$\times$16$\times$16} Monkhorst-Pack wave-vector grid\cite{MP} has been adopted for the integration of the Brillouin zone (BZ) in silicon and GaAs, a {13$\times$13$\times$10} grid was employed for tellurium, while a 16$\times$16$\times$9 grid was used for VSe$_2$.
Phonon frequencies and electron-phonon coupling matrix elements were determined within cDFPT. Phonon frequencies were calculated on  a dense 12$\times$12$\times$12 wavevector grid for photoexcited Si and GaAs, due to the delicate convergence of the phonon modes as a function of the employed wavevector grid, attributed to the presence of the two $quasi$-FSs. Notably, expensive simulations employing a 3456 atoms supercell would have been necessary in order to perform the same calculation within the finite differences approach. The cDFPT formalism described here has been implemented as a modification of the official QE 6.6 release\cite{QE,QE2}.

\section{Conclusions}

In this work, building on top of Refs. \onlinecite{PhysRevB.65.054302,PhysRevLett.82.4340}, we develop a complete
constrained  density functional perturbation theory scheme for structural optimization, calculation of the harmonic 
vibrational properties and electron-phonon interaction  in insulators in the presence of an excited and thermalized electron-hole plasma as the one typically obtained after ultrafast
optical pumping. The method assumes that the photocarriers thermalize faster than the lattice, 
the electron-hole recombination rate is longer than the phonon period and the photocarrier concentration is large enough to screen excitons. 
 We demonstrate its applicability by calculating the evolution of the vibrational spectra as a function of fluence of 
 Te, Si and GaAs. In the case of Te, considering displacive excitations of coherent phonons, we show that allowing for cell relaxation dramatically improves the agreement with experiments for the derivative of the reflectivity peak frequency with respect to fluence (from $14\%$ disagreement in the absence of structural relaxation, to $1\%$ with
 geometrical optimization). In the case of Si and GaAs, we show that even at a fluence corresponding to a PC of $0.1$e$^-_{ph}$/$f.u.$, that is substantially lower that the largest ones achievable in experiments, the phonon spectrum is severely affected with 
 large softenings of optical zone center phonons and emergence of large Kohn-Anomalies. In order to correctly describe these
 anomalies in Si and GaAs, phonon momenta grid as large as $12\times 12\times 12$ are needed. In a finite difference approach 
 this would mean calculating on supercells composed of $3456$ atoms, highlighting the power of our cDFPT approach (and of DFPT approaches in general).
 
  At larger fluences, the Si and GaAs lattice are
 destabilized and imaginary harmonic phonon frequencies emerge, probably a signature of non-thermal melting. However, care is needed as in this regime anharmonic effects could become relevant. As our cDFPT scheme allows a fast access to forces and structural optimization, in the future it could be coupled to the Stochastic Self Consistent Harmonic Approximation\cite{SSCHA_Code} in order to tackle light-induced anharmonicity. 
  
 Finally, as our cDFPT method allows for the study of slowly convergent electronic and vibrational instabilities in photoexcited materials at an affordable computational cost, similarly to what is done by DFPT in ground state studies. It could then be used to screen for
 hidden broken symmetry states and irreversible phase transitions (including non-thermal melting) in materials. 
\section{Acknowledgements}
We acknowledge support from the European Union's Horizon 2020 research and innovation programme Graphene Flagship under grant agreement No 881603. We acknowledge the CINECA award under the ISCRA initiative, for the availability of high performance computing resources and support. We thank Luca Perfetti for useful discussions.

\bibliographystyle{ieeetr}
\bibliography{bibliography}

\end{document}